\documentstyle[multicol,prl,aps,epsfig]{revtex}

\begin{document}
\draft
\def\({\left(}
\def\){\right)}
\def\[{\left[}
\def\]{\right]}
\def\ra{\rangle}
\def\la{\langle}
\def\bege{\begin{equation}}
\def\ende{\end{equation}}
\def\begarr{\begin{eqnarray}}
\def\endarr{\end{eqnarray}}
\def\no{\noindent}
\def\non{\nonumber}
\def\da{\dagger}
\def\e{{\bf{\rm e}}}
\def\i{{\bf{\rm i}}}
\def\tr{{\rm Tr}}
\def\ha{{\hat{a}}}
\def\hb{{\hat{b}}}
\def\tr{{\rm Tr}}
\def\bfr{{\bf {r}}}
\def\bfv{{\bf {v}}}
\def\bfk{{\bf {k}}}
\def\h{{1\over 2}}
\def\hi{\hangindent=10pt}
\def\thefootnote{\fnsymbol{footnote}}
\addtocounter{footnote}{1} \draft

\newcount\hourCount
 \hourCount=\time
 \divide\hourCount by 60

\newcount\minuteCount
 \minuteCount=\time
 \multiply\hourCount by 60
 \global\advance\minuteCount by -\hourCount
 \divide\hourCount by 60

\newcommand{\hour}{\number\hourCount}
\ifnum\minuteCount<10
 \newcommand{\minute}{0\number\minuteCount}
\else
 \newcommand{\minute}{\number\minuteCount}
\fi

\draft
\def\({\left(}
\def\){\right)}
\def\[{\left[}
\def\]{\right]}
\def\ra{\rangle}
\def\la{\langle}
\def\bege{\begin{equation}}
\def\ende{\end{equation}}
\def\begarr{\begin{eqnarray}}
\def\endarr{\end{eqnarray}}
\def\no{\noindent}
\def\non{\nonumber}
\def\da{\dagger}
\def\e{{\bf{\rm e}}}
\def\i{{\bf{\rm i}}}
\def\tr{{\rm Tr}}
\def\ha{{\hat{a}}}
\def\hb{{\hat{b}}}
\def\tr{{\rm Tr}}
\def\bfr{{\bf {r}}}
\def\bfv{{\bf {v}}}
\def\bfk{{\bf {k}}}
\def\h{{1\over 2}}
\def\hi{\hangindent=10pt}
\def\thefootnote{\fnsymbol{footnote}}
\addtocounter{footnote}{1}

\draft

\title{\Large\bf Improving the Efficiency of an Ideal Heat Engine:\\
The Quantum Afterburner}

\author{Marlan O.~Scully}
\address{ Department of Physics and Department of Electrical Engineering, Texas A\&M University, TX~~77843 \\
Max-Planck-Institut f\"{u}r Quantenoptik, D-85748 Garching,
Germany}
\date{\today: \hour: \minute}
\maketitle
\begin{abstract}
By using a laser and maser in tandem, it is possible to obtain laser action in the 
hot exhaust gases involved in heat engine operation.  Such a "quantum afterburner" 
involves the internal quantum states of working gas atoms or molecules as well as the 
techniques of cavity quantum electrodynamics and is therefore in the domain of quantum 
thermodynamics.  As an example, it is shown that Otto cycle engine performance can be 
improved beyond that of the "ideal" Otto heat engine.
\end{abstract}
\begin{multicols}{2}

The laws of thermodynamics \cite{vason} are very useful in telling us how things 
work and what things will never work.  For example, The ideal heat engine is a paradigm 
of modern science and technology.  We read in the textbooks that:\\ 
"it might be supposed that the ideal cycle analysis is too unrealistic to be useful.  In 
fact, this is not so.  Real gas cycles are reasonably close to, although always less 
efficient than, the ideal cycles."

But as technology develops, it behooves us to reexamine thermodynamic dogma.  
The purpose of the present paper is to reconsider the operating limits of ideal heat 
engines in light of recent developments in quantum optics such as cavity QED \cite{haroche}, the 
micromaser \cite{walther}, and quantum coherence effects such as lasing without inversion (LWI) 
\cite{lwi} and cooling via coherent control \cite{cool}.  In particular, we shall show that by extracting 
coherent laser radiation from the "exhaust" gas of a heat engine \cite{sears}, e.g., the Otto cycle 
idealization of the automobile engine.  We here show that it is indeed possible to improve 
on the efficiency of an ideal Otto cycle engine, operating between two fixed temperature 
reservoirs, by adding a quantum afterburner which extracts coherent energy from the hot 
exhaust gases of the heat engine.

In such a quantum Otto engine (QOE) the laser energy is supplied by a thermal 
reservoir in accord with the first law of thermodynamics, and entropy balance is 
maintained as required by the second law \cite{vason}.

In what follows we present a physical picture for and thermodynamic analysis of 
the QOE.  In the conclusion we make contact with related previous work.  In the next 
section we present the quantum engine concept physically as an extension of the 
conventional Otto cycle engine as in Fig.(1).  Then we analyze the QOE, calculating the 
efficiency and entropy flow, etc.  The proposed scheme is simple enough to permit 
reasonably complete analysis; but, hopefully, realistic enough to be convincing.  In the 
conclusion, we examine our results in the context of previous research on the subject.

In order to present the physics behind the QOE, consider Fig.(1) in which the 
working fluid passes through the cycle 1234561.  As mentioned earlier, we extend the 
classical Otto engine to include a laser arrangement which can extract coherent laser 
energy from the internal atomic degrees of freedom.  As depicted in Figs.(1a-e), the QOE 
operates in a closed cycle in the following steps:\\
a.$(1\rightarrow 2)$ The hot gas expands isentropically doing useful ("good") work 
$W_g =C_v (T_1 -T_2)$ where $T_2 =T_1{\mathcal{R}}^{-1}$, $C_v$ is the heat capacity, 
${\mathcal{R}}=(V_1 /V_2)^{(\gamma -1)}$, and $\gamma$ is the ratio of heat capacities at constant 
pressure to constant volume.\\
b.$(2\rightarrow 3)$ Heat $Q_{out} =C_v (T_2 -T_3 )$ is extracted at constant volume by a heat 
exchanger at temperature $T_3$.\\
c.$(3\rightarrow 4)$ Maser-laser cavities are added and energy is extracted from the hot internal 
atomic degrees of freedom by cycling the gas from left to right to left through the 
laser-maser system held at temperature $T_3$, with an entropy decrease $\Delta S_{int}\cong Nk\ln 3$, 
as discussed later.\\
d.$(4\rightarrow 5)$ The gas is then compressed isentropically to volume $V_4 =V_1$, requiring 
waste work $W_w =C_v (T_4 -T_3 )=({\mathcal{R}} -1)T_3$.\\
e.$(5\rightarrow 6)$ The gas is again put in contact with the heat exchanger (at temperature $T_1$) 
and the external-transnational degrees of freedom are heated isochorically to $T_1$ by 
heat energy $Q_{ext}$.\\
f. $(6\rightarrow 1)$ Maser-laser cavities are again added and internal states are heated by an 
amount $q_{in}$ extracted from the hot cavities at temperature $T_1$, completing the cycle.

As a useful simplifying assumption, we consider the external and internal degrees 
of freedom to be decoupled.  Only when the atoms are passing through the maser-laser 
system do they change their internal state.  That is, the atomic states are chosen to be very 
long lived when not in the cavities.  But they are strongly coupled to the radiation field in 
the maser and laser cavities due to the increased density of states of the radiation inside 
the cavity.  Thus, when an atom in the $|b\ra$ state is passed into the maser cavity it quickly 
comes into equilibrium with the thermal radiation in the cavity.  For example in step $3\rightarrow 4$
, after passing through the cold maser, $|b\ra$ state population is determined by the 
Boltzmann factor governed by temperature $T_3$.  And for small enough $T_3$ the $|b\ra$ state is 
effectively depopulated thus providing a population inversion between states $|a\ra$ and $|b\ra$ 
 since the population in state $|a\ra$ is still determined by $T_1$.  This is the basis for lasing 
off the thermal energy of the exhaust gases.

Thus the maser serves as the incoherent ("heat") energy removal mechanism,$q_{in}$, 
which enables the coherent (useful) energy, $w_l$, to be emitted by the laser.  To understand 
the work $w_l$ vs heat $q_m$ aspect of the problem we need only compare the photon statistics 
for the incoherent thermal field in the maser cavity with the coherent laser field.  The 
maser field density matrix is given by\cite{scully}:
$$~~~~~~~~ 
 r_{nn}^{m}=\bar{n}^n_{m}/(\bar{n}_{m}+1)^{n+1}~~~~~~~~~~~~~~~~~~~~~~~~~~~(1a)$$
where $\bar{n}_{m}=1/[\exp(\hbar\nu_{m}/kT_{2}) -1]$, $\hbar\nu_m$ is the 
energy per quantum of the maser field.

The density matrix describing the laser field proceeds from an initial thermal 
state which is largest for small $\bar{n}_l$, to the sharply peaked coherent 
distribution given by$$
r_{n,n}^{(l)}=r_{0,0}\prod^n_{l=1}\left[\frac{A(l+1)}{1+(A/B)(l+1)}-
C(l +1)+\frac{\bar{n}_l}{\bar{n}_l +1}\right] ~~(1b)$$
where A(C) is the linear hain (loss) and B is the nonlinear saturation 
parameter, $\bar{n}_l$ is the average 
number of thermal photons in the laser cavity at temperature $T_2$ with no 
atoms present. 
\setcounter{equation}{7}

Having established the fact that only the laser radiation contributes useful work, we write the 
efficiency of the QOE as:$$
\eta_{qo} = \frac{W_{g}-W_{w}+w_l}{Q_{in}+q_{in}}$$
\setcounter{equation}{1}
and since $q_{in}=w_l +q_m$ we find:
\begin{equation}
\eta_{qo} = \eta_0 +\frac{w_l (1-\eta_0 )-\eta_o q_m}{Q_{in} +w_l +q_m }.
\end{equation}
where the ideal classical Otto engine efficiency is defined as $\eta_{0}=(W_g -W_w )/Q_{in}$.
Taking $W_g$, $W_w$, and $Q_{in}$ as given in the discussion of QOE operation we have the 
alternative expressions $\eta_{0}=1-1/{\mathcal{R}}=1-T_2/T_1$.

In order to determine whether $\eta_{q0}$, Eq.(2), is an improvement over 
$\eta_{0}$ we now turn to the calculation of $w_l$ and $q_m$.  A rigorous 
calculation requires a quantum theory of the laser/maser type analysis and 
this will be given elsewhere.  However it is sufficient for the present 
purposes to apply microscopic energy balance calculations to obtain good 
expressions for the important quantities.

After the atom makes one pass through the maser-laser system, the internal 
density matrix is given by:
\begin{equation}
\rho_{one}(3)=\frac{1}{2}(p_{a}^1 +p_{b}^3 )(\Lambda_a +\Lambda_b )+
(p_{c}^1 +p_{b}^1 -p_{b}^3 )\Lambda_c
\end{equation}
where $\Lambda_a =|\alpha\ra\la\alpha |, \alpha =a, b,$ and $c$, and in the 
notation of Fig.(1), $p_{\alpha}^i$ is the Boltzmann factor given by 
$p_{\alpha}^i =Z^{-1}_i\exp(-\beta_{i}\epsilon_{\alpha})$ where $\beta_i =
1/kT_i$; $T_i$ is the reservoir temperature $T_1$ or $T_2$ and $Z_i =
\sum\exp -(\beta_i\epsilon_\alpha )$.  But an atom will bounce many times back 
and forth through the maser/laser cavities in moving the gas adiabatically 
from right to left.  After many bounces the atom settles down into the mixed 
state:
\begin{equation}
\rho_{many}(3)=p_{b}^3\Lambda_a +p_{b}^3\Lambda_b +(1-2p_{b}^3 )\Lambda_c
\end{equation}

We calculate $q_m$ by noting that $(p_{a}^1 -p_{b}^3 )N$ atoms go from 
$a\rightarrow b\rightarrow c$ and $(p_{b}^1 -p_{b}^3 )N$ atoms go from 
$b\rightarrow c$, and in both cases add energy $\epsilon_b -\epsilon_c$ 
to the maser field. Thus on making the $b\rightarrow c$ transition, the total 
incoherent energy added to the maser field by all N atoms is:
\begin{equation}
q_m =(\epsilon_b -\epsilon_c )N[p_{a}^1 -p_{b}^3 +p_{b}^1 -p_{b}^3 ]
\end{equation}

Likewise $w_l$ is obtained by noting that the number of atoms going from a to b 
with the coherent emission of laser radiation is $N(p_{a}^1 -p_{b}^3 )$ . 
Energy $\epsilon_a -\epsilon_b$ is given up by each atom, and the total 
coherent energy (i.e. useful work) given to the laser field is:
\begin{equation}
W_l=(\epsilon_a -\epsilon_b )N(p_{a}^1 -p_{b}^3 )
\end{equation}

We now use Eqs.(5,6) for $q_m$ and $w_l$ in a form which allows us to 
determine the sign of the efficiency enhancement factor in Eq.(2).  That is, we wish to establish the 
conditions for which:
$$~~~~~~~~~~~~~~(1-\eta_0)w_l >\eta_0 q_m ~~~~~~~~~~~~~~~~~~~~~~~~~~~~~~(7a)$$
We use Eqs.(5,6) and introduce the notation $\epsilon_{\alpha\beta}=
\epsilon_{\alpha} -\epsilon_{\beta}$  to write Eq.(7a) as:
$$\left (\frac{1}{\eta_{0}}-1\right)
\left (\frac{\epsilon_{ac}} {\epsilon_{bc}}-1\right ) > 
1+\frac{p_{b}^1 -p_{b}^3}{p_{a}^1 -p_{b}^3} ~~~~~~~~~~~~~~~~~~(7b)$$
\setcounter{equation}{7}

As an example, we may take $\eta_0 =1/4$, $\epsilon_{ac}/\epsilon_{bc}=11$ so 
that the LHS of Eq.(7b) is 30; furthermore noting that for high enough $T_1$ 
that $p_{\alpha}^1\simeq 1/3$ and so long as $p_{b}^3 < 1/3$ the 
RHS of Eq.(7b) equals 2.  Hence the RHS (maser) factor is an order of 
magnitude less than the LHS (laser) factor in (7a,b), which indeed shows that 
$\eta_{q0} >\eta_0$ as desired. Finally we note that the von Neumann entropy, 
$S=-kNTr\rho\ln\rho$, added in the heating of the internal states to 
temperature $T_1$:
\begin{equation}
S_{int}(6\rightarrow 1)=-kN\sum\left[p_{\alpha}^1\ln p_{\alpha}^1 -
2p_{b}^3\ln p_{b}^3 -p_c\ln p_c\right]
\end{equation}  
where $p_c =1-2p_{a}^3$, is equal and opposite to that removed in the 
$4\rightarrow 4$ maser-laser energy-entropy extraction process.  Hence when 
$T_1$ is high enough and $T_3$ is low enough that $p_c\cong 1$, 
$p_a =p_b =0$ then (8) takes the simple form $S(6\rightarrow 1)\cong kN\ln 3$
 as noted earlier.  We now turn to the relation of the present results to that 
of previous work.

The landmark paper by Ramsey\cite{ramsey}  on negative temperatures in thermodynamics 
and statistical mechanics is a pillar of quantum thermodynamics and is directly relevant 
to the present work.  In his paper he shows that his work is contrary to the Kelvin-Planck 
[10] statement of the second law, which has now been revised to the Kelvin-Planck-
Ramsey statement.  Furthermore in ref\cite{ramsey}  it is noted that: "systems at negative 
temperatures have various novel properties of which one of the most intriguing is that a 
heat engine operating in a closed cycle can be constructed that will produce no other 
effect than the extraction of heat from a negative temperature reservoir with the 
performance of the equivalent amount of work."  But it is also noted that at both positive 
and negative temperatures, cyclic heat engines which produce work have efficiencies less 
than unity, i.e. they absorb more heat than they produce work.

The present study, on the other hand, does not involve negative temperature 
reservoirs.  But it is possible to envision a negative temperature as being associated with 
the $a\rightarrow b$ transition once inversion is produced by the maser interaction, and the present 
work has much in common with that of Ramsey.

The work of Ramsey led to the introduction of the quantum heat engine concept 
by Scovil and Schultz-Dubois.  In their paper\cite{scovil}  entitled, {\it Three level masers as heat 
engines} they conclude that the limiting efficiency of their three level maser engine model 
is that of the Carnot cycle.  And that their work "may be regarded as another formulation 
of the second law of thermodynamics".  In the present paper the atomic states are not to 
be viewed as the engine.  We focus on the different problem of improving the efficiency 
of an ideal heat engine which has a laser-maser system integrated into an Otto cycle 
engine as in Fig. (1).

The present results are an extension of the work by the author and colleagues 
listed in ref\cite{sears} .  In particular, the paper presented at the Dec. 1999 Japanese-American 
Conference on Coherent Control entitled {\it Using External Coherent Control Fields to 
Produce Laser Cooling Without Spontaneous Emission and Sharpen Thermodynamic 
Dogma}\cite{coher} , gave specific examples and direct calculation, based primarily on breaking 
emission symmetry as in lasing without inversion. Thus demonstrating that cooling of 
internal states by external coherent control fields is possible.  There we also showed that 
such coherent schemes allow us to reach absolute zero in a finite number of steps, in 
contrast to usual third "law" of thermodynamics dogma.

It is interesting to compare this work with the paper\cite{gordon} of Kosloff, Geva, and 
Gordon entitled {\it Quantum Refrigerators in the Quest of Absolute Zero} in which they have 
independently arrived at similar conclusions using a similar model.  They have 
established a bound for the maximum cooling rate in the low temperature limit where 
quantum behavior dominates.

In conclusion: we have shown that it is possible, in principal, to improve on the 
efficiency of an ideal Otto cycle engine by extracting laser energy from exhaust gases; 
this is summarized in Fig.(2).  However, as will be presented elsewhere, when a similar 
"lasing-off-exhaust-atoms" scheme is analyzed for a Carnot cycle engine, efficiency is 
not improved.  The present results are in complete accord with the second law.

The author wishes to thank G. Agnolet, H. Bailey, G. Basbas, J. Caton, T. Lalk, S. 
Lloyd, A. Matsko, F. Narducci, N. Nayak, H. Pilloff, N. Ramsey, Y. Rostovtsev, S. 
Scully, and M. S. Zubairy for helpful discussions.  The support of the Office of Naval 
Research, the National Science Foundation, and the Robert A. Welch Foundation is also 
gratefully acknowledged.

\begin{center} \textbf {Figure Captions}
\end{center}
\no Fig. (1) illustrates the steps, 1234561, in cyclic operation of the quantum Otto engine.  
The atomic internal populations are depicted for the three level atom (levels a, b, c) at 
each stage of operation.  A detailed description of operating steps $a\rightarrow f$ is given in the text.

\no Fig. (2) is a temperature (T) – entropy (S) plot for the quantum Otto cycle engine.  Note 
that the entropy is the sum of entropy for the external (kinetic) and internal (quantum) 
degrees of freedom.

\no Fig. (3) depicts the evolution of internal atomic populations for the case in which the 
atom first passes through the maser-laser system; and then (because everything is 
adiabatic and involves long times and many bounces) bounces back and forth through the 
cavities many times.  As discussed in the text, after, a large number of bounces the atom 
settles down into a configuration wherein most of the population is in state c. 
\end{multicols}

 \begin{figure} 
 \center{
\epsfig{file=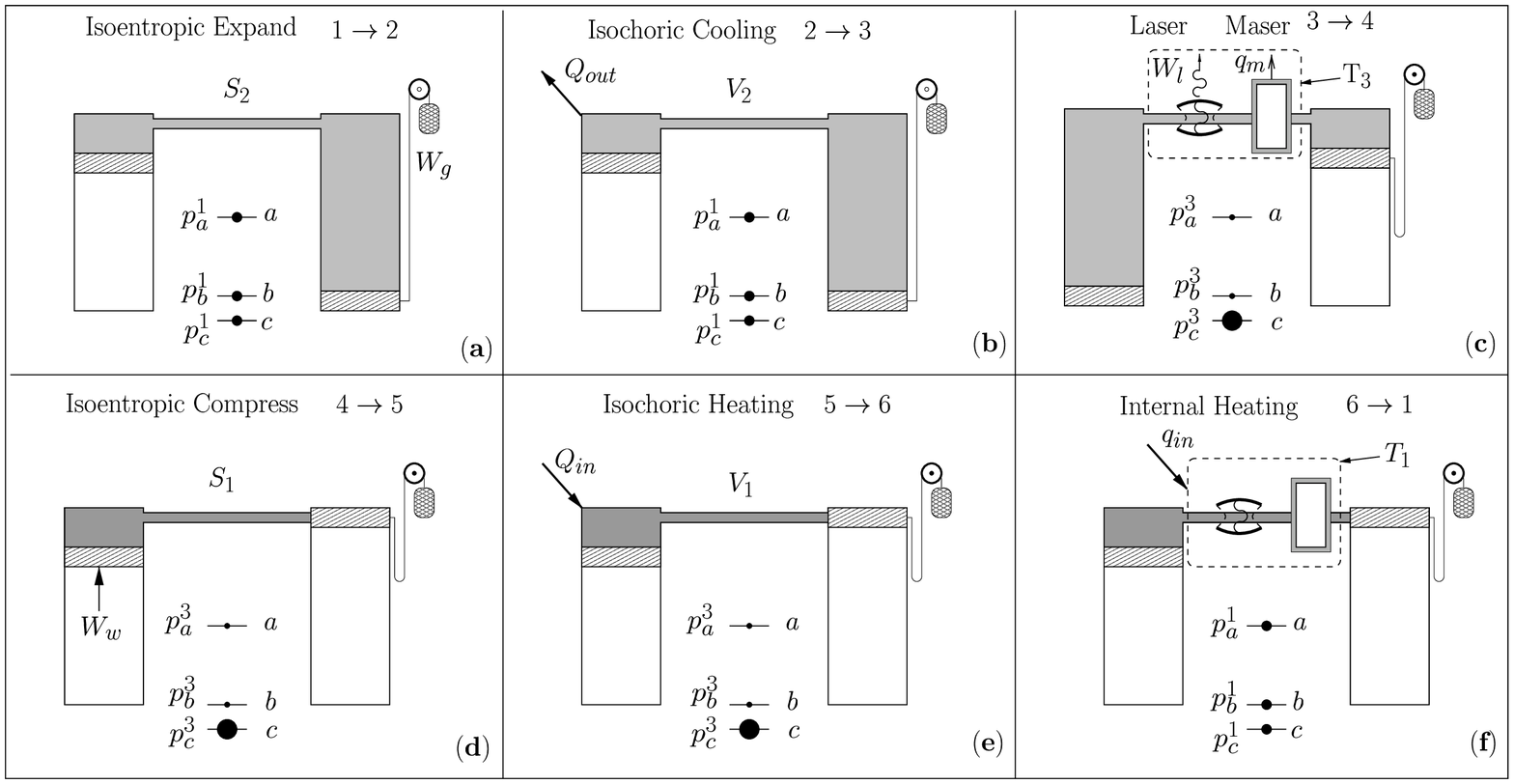, width=18cm, angle=0}
}
 \caption{}
 \end{figure}

 \begin{figure} 
 \center{
\epsfig{file=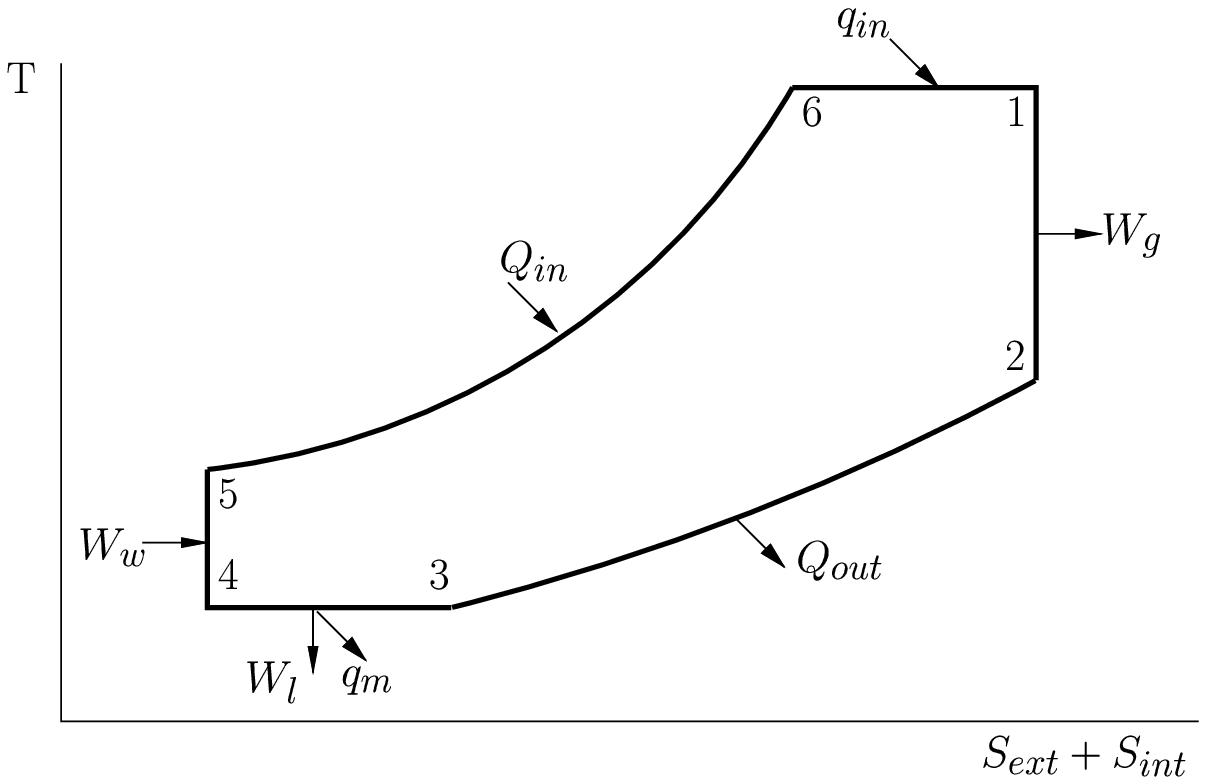, width=10cm, angle=0}
}
 \caption{}
 \end{figure}

 \begin{figure} 
 \center{
\epsfig{file=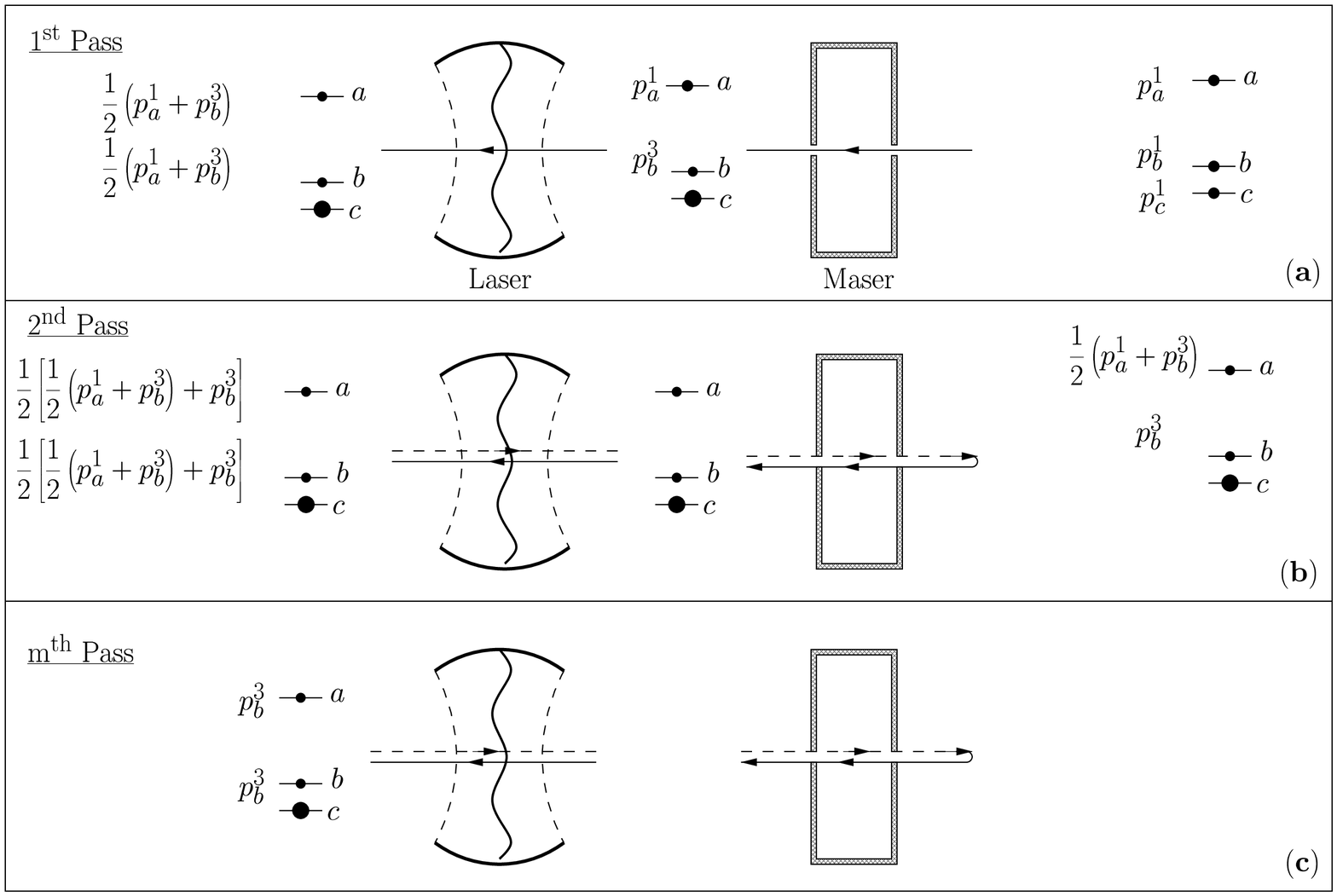, width=18cm, angle=0}
}
 \caption{}
 \end{figure}

\end{document}